\documentclass{kapproc} 
 
\usepackage{procps}  
 
\usepackage[dvips]{graphicx} 
 
\upperandlowercase 
 
\let\footnote\savefootnote 
 
\kluwerbib 
 
\usepackage{epsfig}

\begin{document} 
 
\articletitle{Isolated neutron stars: An astrophysical perspective} 
 
 
 
 
\author{Sergei Popov,\altaffilmark{1,2} and Roberto Turolla\altaffilmark{1} 
} 
 
\altaffiltext{1}{University of Padova\\ 
via Marzolo 8, 35131\\ 
Padova, Italy} 
\email{turolla@pd.infn.it, popov@pd.infn.it} 
 
\altaffiltext{2}{Sternberg Astronomical Institute\\ 
Universitetski pr. 13, 119992\\
Moscow, Russia} 
\email{polar@sai.msu.ru} 
 
\begin{abstract} 
We briefly review selected results in astrophysics of neutron stars (NSs)
obtained during the last two years, focusing on isolated 
radioquiet objects.
We discuss in some details population synthesis of close-by isolated NSs, 
spectra of INSs, detection of spectral features in these sources
(including cyclotron features), recent results on velocity distribution of
NSs and accretion onto INSs.
\end{abstract} 
 
 
\section*{Introduction} 

Probably neutron stars (NSs) are the most interesting astronomical objects
from the physical point of view. They provide a variety of different
{\it extreme} phenomena: magnetic field over the QED limit,
supranuclear density, superfluidity, superconductivity,
exotic matter states, etc.

\begin{figure}[t]
\epsfig{file=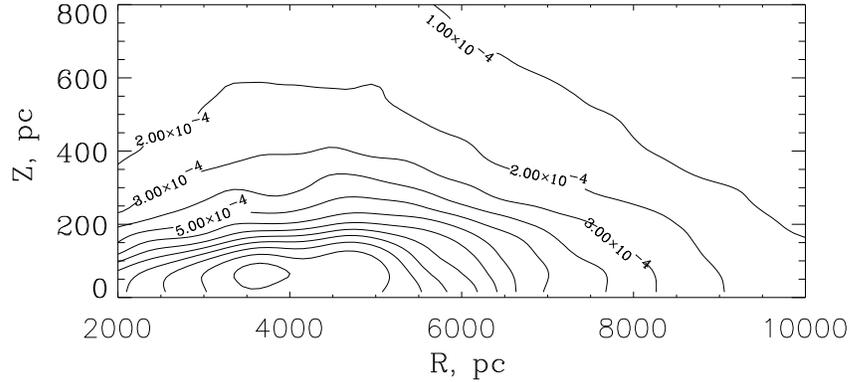, width=\textwidth}
\caption{Spatial distribution of NSs in the Galaxy.
The data was calculated by a Monte-Carlo simulation.
Kick velocity was assumed following Arzoumanian et al. (2002).
NSs were born in the thin disk with semithickness 75 pc.
No NS born inside $R=2$~kpc and outside $R=16$~kpc were taken into account.
NS formation rate was assumed to be constant in time
and proportional to the square of the ISM density at the birthplace. 
Results were normalized to have in total
$5\times 10^8$ NSs born in the described region. 
Density contours are shown with a step
$0.0001$~pc$^{-3}$. At the solar distance from the center close to the
galactic plane the NS density is about $2.8\cdot 10^{-4}$~pc$^{-3}$.
From Popov et al. (2003a).}
\label{dens}
\end{figure}

There are about $10^8$~--~$10^9$ NSs in the Galaxy 
and their local density is about $3\times 10^{-4}$~pc$^{-3}$ (see
fig.~\ref{dens}).
\footnote{Here and below we will
not distinguish between NSs, quark stars, hybrid stars etc. unless
explicitely stated.} At present
only a small fraction (about 2000 sources) of this large population 
 is observed as isolated objects
of different nature and as accreting objects or millisecond radio pulsars
in binaries.
Young NSs can be observed for several million years dissipating their
rotational, thermal or/and magnetic energy.
Most of old NSs are dim objects without significant internal sources of energy.
If a NS looses its strong magnetic field on a time scale
$<10^8$--$10^9$~yrs then it can be resurrected by accretion from 
the interstellar medium (ISM)
or from a binary companion (small number of old NSs  can be 
spin-up by disc accretion and appear as millisecond radio pulsars). 
However, as we will discuss later on, there is no much hope
that a significant number of isolated NSs can be bright accretors, so most 
of NSs are unobservable.
 
The main parameters which determine the astrophysical appearance of NSs
are:
\begin{itemize}
\item Spin period, $p$
\item Magnetic field, $B$
\item Mass, $M$
\item Spatial velocity, $v$
\item Surface temperature, $T$
\item Angle between spin and magnetic axis, $\alpha$

\end{itemize}
Parameters of the surrounding medium (interstellar medium or matter
from the binary companion) are also important.

Here we will focus on isolated NSs (INSs).
At present the following types of these sources are observed:

1. Radio pulsars (PSRs)

2. Anomalous X-ray pulsars (AXPs) 

3. Soft gamma repeaters (SGRs)

4. Compact central objects in supernova remnants (CCOs in SNRs)

5. Geminga and geminga-like object(s)

6. The ``Magnificent seven'' --- seven dim ROSAT sources

INSs may lurke within unidentified EGRET and
ROSAT sources and, possibly, among dim X-ray sources observed by
XMM-Newton and Chandra
in globular cluster (see Pfahl, Rappaport
2001) or in the galactic center (see Muno et al. 2003).
Except PSRs all others are more or less radioquiet (which, however, does not
mean that they are necessarily radio silent). 

Astronomy is the only purely observational natural science.
All the information we have come through the electromagnetic emission
from celestial bodies --- no direct experiments are
possible (of course except some rare cases in the Solar system). 
This is why progress in astronomy is necessarily
connected with new observational facilities.
 Astrophysics of NSs is a quickly growing field. 
New space observatories (especially XMM-Newton and Chandra) give
us an opportunity to obtain increadible
spatial and spectral resolution in the X-ray band. New radio surveys of
PSRs doubled the number of know objects of this type over the last few
years. 
Data from optical and IR telescopes foster new discoveries in NS
astrophysics.
The huge flow of new observational data stimulated theoretical studies.
Here we briefly review recent results in astrophysics of NSs which can be of
interest for participants of this conference -- mainly physicists working on
quark stars and related subjects.
In the  next section we just give a list of them 
(we do not try to summarize results
directly connected to quark stars since they are presented in other
contribution in this proceedings). 
Then in the following
sections we comment on some of them in more details, focusing on INSs
and paying more attention to the results connected with our own
research. 

\begin{table} 
\caption{Local ($D<1$ kpc) population of young (age $<4.25$ Myrs) 
isolated neutron stars\vspace*{1pt}} 
{\footnotesize 
\begin{tabular}{|l||c|c|c|c|c|c|} 
\hline 
 
{} & {} & {} &{} &{} &{} &{}\\[-1.5ex] 
Source name & Period & CR$^a$ & $\dot P$ & D & Age$^b$ & Refs 
\\[1ex] 
            &   s     & cts/s & $10^{-15}$ s/s& kpc   & Myrs     & 
\\[1ex] 
\hline 
RINSs & & & & & & \\[1ex] 
RX J1856.5-3754            &  ---  & 3.64  &  ---&0.117$^e$&$\sim0.5$& 
[1,2]\\[1ex] 
RX J0720.4-3125                 & 8.37 & 1.69  &$\sim 30-60$& ---&---& 
[1,3]\\[1ex] 
RX J1308.6+2127 & 10.3 & 0.29  & --- & --- & --- & 
[1.4] \\[1ex] 
RX J1605.3+3249       &  ---  & 0.88  & --- & --- & --- & 
[1]\\[1ex] 
RX J0806.4-4123                 &  11.37  & 0.38  & --- & --- & --- & 
[1,5]\\[1ex] 
RX J0420.0-5022                 &  3.45   & 0.11  & --- & ---   & --- & 
[1,11]\\[1ex] 
RX J2143.7+0654 &  ---    & 0.18  & --- & ---  & --- & 
[6]\\[1ex] 
\hline 
 Geminga type& & & & & & \\[1ex] 
PSR B0633+17            & 0.237 & 0.54$^d$ &10.97&0.16$^e$&0.34& 
[7]\\[1ex] 
3EG J1835+5918 & ---   & 0.015    & --- & ---  &  --- & 
[8]\\[1ex] 
\hline 
 Thermally emitting PSRs & & & & & & \\[1ex] 
PSR B0833-45    & 0.089 & 3.4$^d$  & 124.88 & 0.294$^e$ & 
0.01& 
[7,9,10]\\[1ex] 
PSR B0656+14          & 0.385 & 1.92$^d$ &  55.01 & 0.762$^f$ & 0.11 
& 
[7,10]\\ [1ex] 
PSR B1055-52          & 0.197 & 0.35$^d$ &   5.83 & $\sim 1^c$ & 
0.54& 
[7,10]\\ [1ex] 
PSR B1929+10          & 0.227 & 0.012$^d$& 1.16 &  0.33$^e$  & 3.1& 
[7,10]\\[1ex] 
\hline 
Other PSRs & & & & & & \\[1ex] 
PSR J0056+4756  & 0.472 & --- & 3.57 &  0.998$^f$  & 2.1& 
[10]\\[1ex] 
PSR J0454+5543  & 0.341 & --- & 2.37 &  0.793$^f$  & 2.3& 
[10]\\[1ex] 
PSR J1918+1541  & 0.371 & --- & 2.54 &  0.684$^f$  & 2.3& 
[10]\\[1ex] 
PSR J2048-1616  & 1.962 & --- & 10.96&  0.639$^f$  & 2.8& 
[10]\\[1ex] 
PSR J1848-1952  & 4.308 & --- & 23.31&  0.956$^f$  & 2.9& 
[10]\\[1ex] 
PSR J0837+0610  & 1.274 & --- & 6.8  &  0.722$^f$  & 3.0& 
[10]\\[1ex] 
PSR J1908+0734  & 0.212 & --- & 0.82 &  0.584$^f$  & 4.1& 
[10]\\[1ex] 
\hline 
\multicolumn{7}{l}{ 
$^a$) ROSAT PSPC count rate; 
$^b$) Ages for pulsars are estimated as $P/(2\dot P)$,}\\ 
\multicolumn{7}{l}{ 
      for RX J1856 the estimate of its age comes from kinematical 
      considerations.}\\ 
\multicolumn{7}{l}{ 
$^c$) Distance to PSR B1055-52 is uncertain ($\sim$ 0.9-1.5 kpc)}\\ 
\multicolumn{7}{l}{ 
$^d$) Total count rate (blackbody + non-thermal)}\\ 
\multicolumn{7}{l}{ 
$^e$) Distances determined through parallactic measurements}\\ 
\multicolumn{7}{l}{ 
$^f$) Distances determined with dispersion measure}\\ 
\multicolumn{7}{l}{ 
[1] Treves et al. (2000) ; [2] Kaplan et al. (2002); [3] Zane et al. (2002);}\\ 
\multicolumn{7}{l}{ 
[4] Hambaryan et al. (2001); [5] Haberl, Zavlin (2002); [6] Zampieri et al. 
(2001);}\\ 
\multicolumn{7}{l}{ 
[7] Becker, Trumper (1997); [8] Mirabal, Halpern (2001); [9] Pavlov et al. 
2001;}\\ 
\multicolumn{7}{l}{ 
[10] ATNF Pulsar Catalogue (see Hobbs et al. 2003); 
[11] Haberl et al. (2004, in
prep.)}\\ 
\hline 
\end{tabular} } 
\vspace*{-13pt} 
\end{table} 

 
\section{What's new}  

In this section 
we give a list of new important discoveries in obervational and in
theoretical astrophysics of NSs. In the observational part of the list
we usually give  objects names, determined parameters and reference to the
original paper.
In the theoretical part we sometimes just name the topic
of research and give references to original papers or/and reviews
on that topic.

\subsection{Observations}
 
1. Magnetic field determination from cyclotron features

\begin{figure}[t]
\epsfig{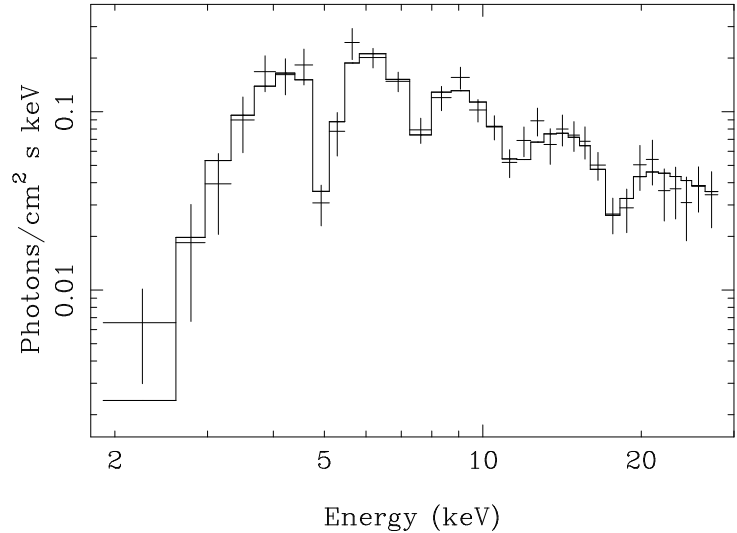}
\caption{Spectrum of SGR 1806-20. 
Observed by RXTE.
From Ibrahim et al. (2002).}
\label{ibra}
\end{figure}

\begin{figure}[t] 
\epsfig{file=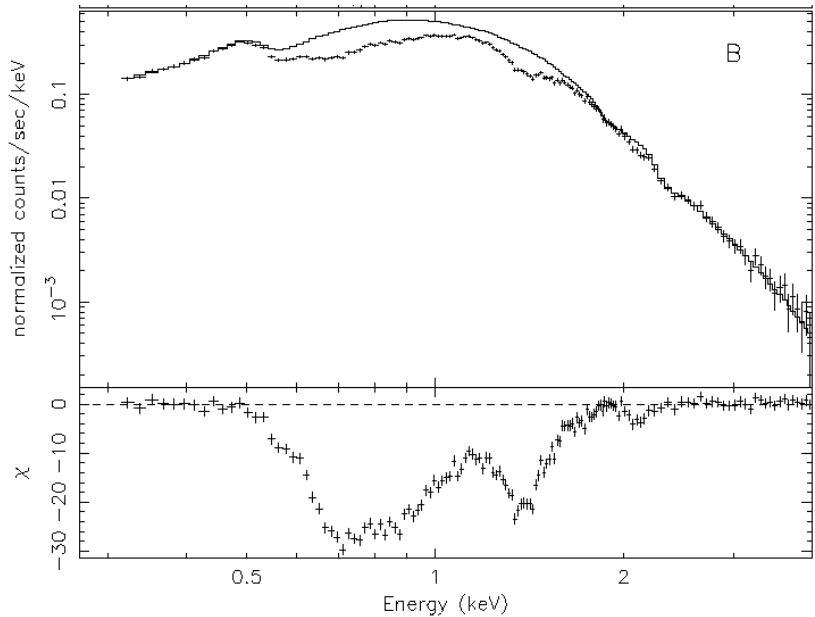, width=\textwidth}
\caption{Spectrum of 1E 1207.4-5209 collected by the MOS camera on the EPIC
instrument of XMM-Newton. Also the best fit continuum
and residuals are shown. From Bignami et al. (2003).}
\label{bign}
\end{figure}

\begin{itemize}
\item SGR 1806-20. $B\sim 10^{15}$~G 
(if it is a proton cyclotron resonance feature) (Ibrahim et al. 2002).
See fig.~\ref{ibra}.
\item AXP 1RXS J170849-400910. 
$B \sim 9\cdot 10^{11}$~G (electron resonance) 
or $1.6 \cdot 10^{15}$~G (proton resonance) (Rea et al. 2003). 
\item AXP 1E 1048-5937. $B\sim 1.2\cdot 10^{12}$~G (electron resonance) or
$B\sim 2.4\cdot 10^{15}$~G (proton resonance) (Gavriil et al. 2002,
2003). Not a very strong feature. Not consistent with spin-down.
\item 1E 1207.4-5209. CCO in SNR. $B\sim 8\cdot 10^{10}$~G 
(electron resonance) or 
$\sim 1.6\cdot 10^{14}$~G (proton resonance)
(Bignami et al. 2003). From spin-down measurements the field estimate is
$B\sim (2-3)\cdot 10^{13}$~G (see fig.~\ref{bign}).
\item RBS 1223. "Magnificent seven". $B\sim (2-6) \cdot 10^{13}$~G 
(proton resonance) (Haberl
et al. 2003).
\end{itemize}

\begin{figure}[t] 
\epsfig{file=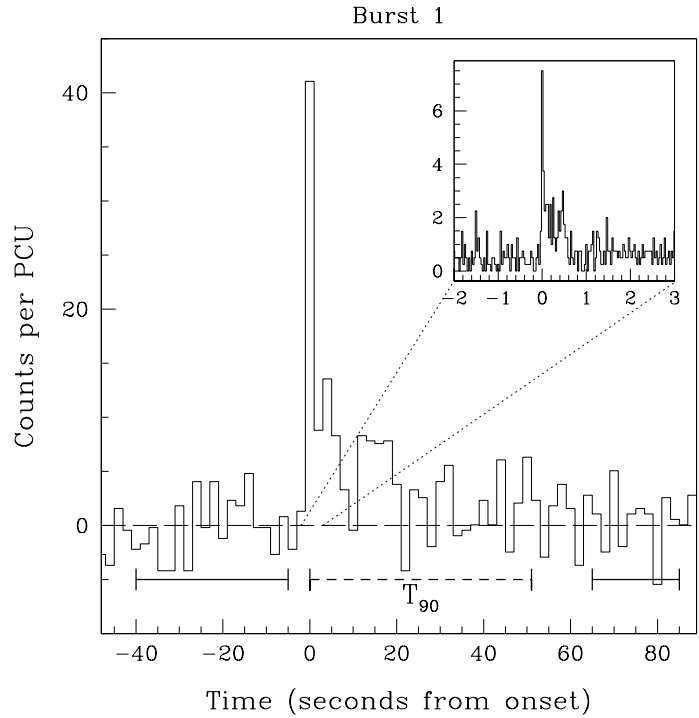, width=0.8\textwidth}
\caption{AXP 1E 1048-5937 burst. 
Observed by RXTE. From Gavriil et al. (2002).}
\label{1048}
\end{figure}

2. Relations between AXPs and SGRs

Observations of X-ray bursts from AXPs
which are very similar to the ones from SGRs.

\begin{itemize}
\item AXP 1E1048-5937 (Gavriil et al. 2002, 2003). See fig.~\ref{1048}.
\item AXP 1E 2259+586 (Kaspi, Gavriil 2002).
\end{itemize}

3. More radiopulsars.

\begin{itemize}
\item On-line ATNF catalogue. 1300 PSRs (Hobbs et al. 2003).
\item  Parkes survey.  $>800$ new PSRs 
(Kramer et al. 2003, Morris et al. 2002).
\item Pulsar with magnetar parameters:
$p=6.7$~s, $B\approx 9.4\cdot 10^{13}$~G (McLaughlin et al. 2003).
\item A new double NS system (Burgay et al. 2003).
\end{itemize}

4.  NS initial velocity distribution

\begin{itemize}
\item New PSRs proper motions (Brisken et al. 2003).
\item New model for the galactic distribution of free electrons (Cordes, Lazio
2002).
\item Bimodal initial velocity distribution with two 
maxwellian components: $\sigma_1$=90~km~s$^{-1}$ and 
$\sigma_2$=500~km~s$^{-1}$
 (Arzoumanian et al. 2002).
\end{itemize}

5. Proper motions of radioquiet INSs 

\begin{itemize}
\item RX 1856.5-3754. $d=117$ pc, $v_T=185$~km~s$^{-1}$.
Excludes accretion from the ISM
 (Walter, Lattimer 2002; Kaplan et al. 2002).
Distance to this source is still uncertain.
\item RX 0720.4-3125. $v_T=50 (d/100 {\rm pc})$~km~s$^{-1}$.
Excludes accretion from the ISM  
(Motch et al. 2003).
\end{itemize}

6. Discovery of a new Geminga-like object

\begin{itemize}
\item 3EG 1835+5918. EGRET source  (Halpern et al. 2002).
\end{itemize}

7. $\dot p$ for radioquiet INS.

\begin{itemize}
\item RX 0720.4-3125. 
$\dot p\sim (3-6) \cdot 10^{-14}$ (Zane et al. 2002).
\item Kes 75. $p=0.325$~s, $\dot p=7.1 \cdot 10^{-12}$ 
(Mereghetti et al. 2002).
\item G296.5+10. $p=0.424$~s, $\dot p=(0.7-3) \cdot 10^{-14}$ (Pavlov et al.
2002).
\item Non-constant $\dot p$ for 1E 1207.d-5209 (Zavlin et al. 2003). 
\end{itemize}

8. IR radiation from AXPs

\begin{itemize}
\item 1E 2259+586 (Hulleman et al. 2001).
\item 1E 1048.1-5937 (Wang, Chakrabarty 2002).\\
Variability (Israel et al. 2002).
\item 1RXS J1708.9-400910 (Israel et al. 2003).
\end{itemize}

9. Pulsars jets and toruses

\begin{itemize}
\item Variable jet of the Vela pulsar (Pavlov et al. 2003)
\item Alignment between spin axis
and spatial velocity for Crab and Vela pulsars (see for example Lai et al.
2001)
\end{itemize}

10. NS masses

\begin{itemize}
\item Vela X-1. $M\approx 2\, M_{\odot}$ (Quaintrell et al. 2003).
\item PSR J0751+1807. $M\approx1.6-1.28\, M_{\odot}$ (Nice, Splaver 2003).
\end{itemize}

11. Gravitationaly redshifted line from an accreting NS

\begin{itemize}
\item EXO 0748-676. z=0.35 (Cottam et al. 2002).
\end{itemize}

\subsection{Theory}

1. Spectra of strongly magnetized NSs 

\begin{itemize}
\item Lines for high magnetic field 
(Zane et al. 2001; Ho, Lai 2001, 2003; Ozel 2001).
\item Bare neutron and quark star emission (Turolla et al. 2004).
\item Atmospheres and opacities for high magnetic fields
(Potekhin, Chabrier et al.).
\end{itemize} 

2. Gould Belt in population synthesis calculations

\begin{itemize}
\item Population synthesis of EGRET sources (Grenier 2003).
\item Population synthesis of young cooling INS (Popov et al. 2003a).
\end{itemize}

3. SN explosions (see the contribution by  Stephan Rosswog in this volume
for more details)

\begin{itemize}
\item 3-Dimensional core-collapse (Fryer, Warren 2003).
\item Nucleosynthesis, collapse dynamics (Woosley et al. 2002).
\item Jets, GRB connection, X-ray flashes, HETE-2 data (Lamb et al. 2003).
\end{itemize}

4. Accretion and spin evolution

\begin{itemize}
\item CDAF -- convection dominated accretion flows (Igumenschev et al. 2002,
2003)
\item Low angular momentum accretion (Proga, Begelman 2003).
\item Accretion onto INSs (Toropina et al., Romanova et al. 2003).
\item Propeller regime for INSs (Ikhsanov 2003).
\end{itemize}

5. Cooling curves 

\begin{itemize}
\item Impact of superfluidity on cooling (Kaminker et al. 2003,
Tsuruta et al. 2002).
\item Cooling curves for quark stars (see  contributions by 
Grigorian et al. and others in this procedings)
\end{itemize} 

6. Discussion on models of fossil discs around INSs

\begin{itemize}
\item Discs can explain AXPs and other types of INS  (Alpar 2003).
\item Discs can't explain it (Francischelli, Wijers 2002).
\item General picture of pulsars with jets and disks (Blackman, Perna 2003).
\end{itemize} 

7. Electrodynamics of magnetars 

\begin{itemize}
\item SGR phenomena due to magnetospheric activity (Thompson et al. 2002).
\end{itemize}

\section{Discussion}

Here we discuss some results in more details focusing on topics of our
personal interest.

\subsection{Spectral features and magnetic field determination}

\begin{figure}[t]
\epsfig{file=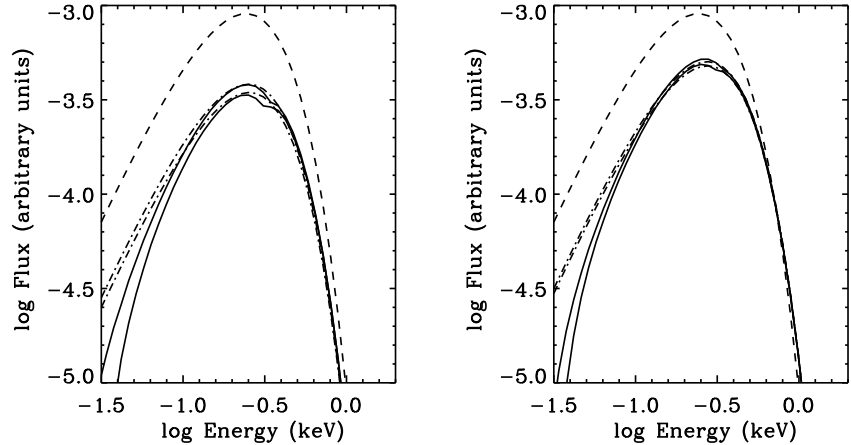, width=\textwidth}
\caption{Model spectra of a naked neutron  star. 
The emitted spectrum with electron-phonon damping
accounted for and $T_{surf}=10^6$~K. Left
panel: uniform surface temperature; right panel: meridional
temperature variation. The dashed line is the blackbody at
$T_{surf}$ and the dash-dotted line the blackbody which best-fits
the calculated spectrum in the 0.1--2 keV range. The two models
shown in each panel are computed for a dipole field
$B_p=5\times 10^{13}$~G
(upper solid curve) and $B_p=3\times 10^{13}$~G (lower solid curve).
Spectra are at the star surface and no red-shift correction has
been applied.
From Turolla, Zane and Drake (2004).}
\label{fig:tur1}
\end{figure}


Investigations on the emission properties of INSs
started quite a long time ago, mainly in connection with
the X-ray appearance of PSRs. In the '70s it was common wisdom
that the radiation emitted by INSs come directly from their solid
crust and is very close to a blackbody. Lenzen and Trumper
(1978) and Brinkmann (1980) were the first to address in detail
the issue of the spectral distribution of INS surface emission.
Their main result was that the star emissivity is strongly
suppressed below the electron plasma frequency which, for the
case of X-ray pulsars, is at about 1 keV.

Later on the role played by the thin atmosphere covering the star
crust in shaping the emergent spectrum started to be appreciated.
Romani (1987) investigated the properties of fully-ionized, H
atmospheres around unmagnetized, cooling NSs and showed that the
spectrum is harder than a blackbody at the star effective
temperature. Spectra from magnetized NSs were studied, under
similar hypotheses, by the St. Petersburg group in a series of
papers (Shibanov et. al 1992; Pavlov et al. 1994; Zavlin et al.
1995). Because the opacity is higher in the presence of a strong
magnetic field ($B\sim 10^{12}$~G or larger), magnetic spectra tend to
be more blackbody-like. In these investigations, the focus was on
middle-aged NSs with $T_{eff}\approx 10^5-10^6$~K and typical
fields $\approx 10^{12}$--$10^{13}$~G. Under such conditions the
bulk of the emission is the soft X-ray band ($\approx
0.1$--1~keV) while the electron cyclotron line at
$\hbar\omega_{c,e}\sim 11.6 (B/10^{12}\, {\rm G})$~keV falls in
the tens of keVs range. For this reason no detailed modeling of
the line was attempted. An (approximated) treatment of the proton
cyclotron line was included, although its energy,
$\hbar\omega_{c,p}\sim 6.3 (B/10^{12}\, {\rm G})$~eV, falls in the
optical/UV region for pulsar-like fields. Being related to a
resonant behaviour of the opacity for the extraordinary mode, the
proton cyclotron line appears as an absorption feature in the
spectrum. Atmospheres comprised of heavy elements (Fe) were
studied by Rajagopal et al. (1997); the emergent
spectra exhibit a variety of emission/absorption features produced
by atomic transitions. Such models, however, suffer from our lack of
knowledge on ionization state and opacities of metals in a strong
magnetic field.

Until quite recently, model atmosphere calculations were
restricted to fields not exceeding a few $10^{13}$~G.
Ultra-magnetized NSs has been long suspected to exist in SGRs. 
It was only in the late '90s that the
positive detection of large spin-down rates in SGR 1806-20 and
SGR 1900+14 (Kouveliotou et al. 1998; 1999) provided decisive
evidence in favor of the magnetar scenario. At about the same time
spin-down measures supported the magnetar nature of the AXPs. 
In nearly all AXPs and in at
least in one SGR, a thermal component was clearly detected in the
X-ray spectrum.  This prompted renewed interest in the study of
the thermal emission from NSs with surface fields in the
$10^{14}$--$10^{15}$~G range. The main goal was to identify
possible signatures of the super-strong magnetic field which
could provide an unambiguous proof of the existence of magnetars.
The proton cyclotron resonance, being in the keV range for a
magnetar, is an ideal candidate for this. Zane et al. (2001) were
the first to construct model atmospheres for $B\sim
10^{14}$--$10^{15}$~G and luminosities appropriate to
SGRs/AXPs. They considered completely ionized, pure H atmospheres
in radiative equilibrium and solved the transfer problem in a
magnetized medium in the normal modes approximation and planar
symmetry. Computed spectra are blackbody-like and show a
relatively broad absorption line at $\hbar\omega_{c,p}$ with an
equivalent width ${\rm EW}\approx 100$~eV, within the detection
capabilities of  Chandra and XMM-Newton. This issue
was further addressed by Ho and Lai in a recent series of papers
(Ho, Lai 2001, 2002; Lai, Ho 2003). Their approach differs
from that used by Zane et al. (2001) in the treatment of vacuum
polarization and mode conversion. This affects the line
properties in a non-negligible way. In the first paper no
adiabatic mode conversion was included and the line EW they found
is larger than that predicted by Zane et al. Accounting for
adiabatic mode conversion produces a depression in the continuum
at energies close to the cyclotron energy, thus reducing the line
EW to values somewhat smaller than those of Zane et al. Spectra
from ultra-magnetized NSs were also computed by Ozel (2001)
who, however, did not account for the proton contributions.

While a proper solution of transfer problem in media at $B\gg
B_{QED}\simeq 4.4\times 10^{13}$~G has necessary to wait for a
description in terms of the Stokes parameters, the search for the
proton cyclotron feature in the spectra of AXPs and SGRs begun.
Up to now no evidence for the proton line has been found in the
thermal components of SGRs and AXPs, although these
observations can not be regarded as conclusive yet. Quite recently
a cyclotron absorption feature in the spectrum of SGR 1806-20 in
outburst has been reported by Ibrahim et al. (2002) and further
confirmed by the same group in several other events from the same
source (Ibrahim et al. 2003). The line parameters are similar to
those predicted by Zane et al. (2001), even if the line in this
case is not superimposed to a thermal continuum and is not
expected to originate from the star cooling surface. The line
energy ($\simeq 5$~keV) implies a field strength $B\sim 10^{15}$~G
in excellent agreement with the spin-down measure.

Model spectra from standard cooling NSs proved successful in
fitting X-ray data for a number of sources and in some cases
solved the apparent discrepancy between the star age as derived
from the temperature and from the $\dot P/2P$ measure.
However, model atmospheres seem to be of
no avail in interpreting the multiwavelength spectral energy distribution
(SED) of the seven 
ROSAT INSs. The X-ray spectrum of the most luminous source RX
J1856.5-3754 is convincingly featureless and shows, possibly,
only slight broadband deviations from a blackbody (Drake et al.
2002; Burwitz et al. 2003). The situation is more uncertain for
the fainter sources, and the possible presence of a
(phase-dependent) broad feature at 200--300~eV has been reported
very recently in RBS 1223 (Haberl et al. 2003). In all the cases
in which an optical counterpart has been identified, the optical
flux lies a factor $\approx 5-10$ above the Rayleigh-Jeans tail of
the blackbody which best-fits the X-rays (Kaplan et al. 2003).

The small radiation radius implied by the distance ($\sim 120$~pc,
but this value is still under debate) led to the suggestion that
RX J1856.5-3754 may host a quark star (Drake et al. 2002; Xu 2002, 2003).
Other, more conventional explanations are well possible. Pons et
al. (2002) and Braje and Romani (2002) suggested a scenario in
which the X-rays come from a hotter region close to the poles,
while the reminder of the star surface is at lower temperature and
produces the optical/UV flux. While this picture is appealing and
still consistent with the lack of pulsations (pulsed fraction
$<1.3\%$ see Haberl et al. 2003), no explanation is offered for the
formation of a pure blackbody spectrum in an object which should
conceivably be covered by an optically thick atmosphere. Very
recently Turolla et al. (2004) considered the possibility
that RX J1856.5-3754 is a bare NS, that is to say no atmosphere
sits on the top of its crust. Lai and Salpeter (1997; see also Lai
2000) have shown that for low surface temperatures and high
enough magnetic fields, the gas in the atmosphere undergoes a
phase transition which turns it into a solid. While the onset of
such a transition appears unlikely for an H atmosphere, it might
be possible for a Fe composition for the temperature of  RX
J1856.5-3754 ($T_{BB}\sim 60$~eV) and $B> 3-5\times 10^{13}$~G.
Turolla et al. computed the spectrum emitted by the bare
Fe surface including electron-phonon damping in the highly
degenerate crust, and found that it is close to depressed
blackbody. If indeed RX J1856.5-3754 is bare NS, and keeping in
mind that their results depend on the assumed properties of the
crust-vacuum interface, the optical/UV emission may be due to a
thin H layer which cover the star and is optically thick up to
energies $\approx 10-100$~eV. The Rayleigh-Jeans optical/UV
emission is at the star surface temperature, and the optical
excess with respect to the X-ray spectrum arises because the
latter is depressed.

\subsection{Velocity distribution}

The number of PSRs with known transverse velocities is continuously growing.
New velocity determinations are based on a new model of galactic distribution
of free electrons (Cordes, Lazio 2002). Unlike situation 10 years ago, when
updated data on free electrons distribution leaded to nearly 
doubling of distances (and, correspondently, transverse velocities), 
results of Cordes and Lazio brought
serious corrections only for distant PSRs. 

In last two years a new initial velocity distribution of NSs became
standard. It is a bimodal distribution with peaks at $\sim 130$~km~s$^{-1}$
and  $\sim 710$~km~s$^{-1}$ (Arzoumanian et al. 2002).
Contribution of low and high velocity populations is nearly equal.
Brisken et al. (2003) confirm this type of distribution, however they give
arguments for smaller fraction of low velocity NSs (about 20\%).

The nature of this bimodality is unknown, and recently several
papers appeared where authors suggested (or modified) 
different kick mechanisms.
There are three main mechanism for a natal kick (see for example Lai et al.
2001).
The first is a hydrodynamical one. In many models 
of that mechanism NSs do not recieve
kicks higher than $\sim (100-200)$~km~s$^{-1}$ due to it
(Burrows et al. 2003). The second one
is a modification of an electromagnetic rocket mechanism (see Huang et al.
2003).
In this scenario  velocity is dependent on the initial spin rate
($v\propto p^{-2}$).
It can provide high velocities if the initial spin period of a NS is about
1~ms (probably quark stars can spin faster than NSs, 
so for them this mechanism can be more effective). 
The third mechanism is connected with instabilities in a
newborn NS which lead to it fragmentation into two stars followed by an
explosion of the lightest one (see Colpi, Wasserman 2002). 
Also one should have in mind disruption of
high-mass binaries, so that a newformed compact objects recieve significant
spatial velocity due to orbital motion even without any natal kick
(Iben, Tutukov 1996). 
However, this mechanism can not provide enough number of high
velocity NSs to explain the second peak of the distribution.

In connection with quark stars 
one can speculate, that additional energy due to deconfinement
can lead to additional kick, so among high velocity compact objects
the fraction of quark stars can be higher. For example, if the delayed 
deconfinement proposed by Berezhiani et al. (2003) is operating
(see the contribution by Bombaci in this proceedings), 
then quark stars can obtain additional (second) kick.
Note, that two the most studied NSs -- Crab and Vela  
(which both show spin-velocity alignment, glitches
and other particular properties) -- belong to low velocity population. 
Also most of compact objects in binaries in the scenario with delayed
deconfinement
should belong to low velocity normal NSs as far as otherwise there is a high
probability of system disruption.

\subsection{Young close-by NSs and the Gould Belt} 

The Gould Belt  is a structure
consisting of clusters of massive stars. The Sun is situated not far from
the center of  that disc-like structure. 
The Gould Belt radius is about 300 pc. It is inclined at 18$^\circ$ respect
to the galactic plane.

Due to the presence of the Belt the rate of SN
around us (say in few hundred parsecs) during last several tens of million
years is higher, than it is in an average place at a solar distance from the
galactic center. Because of that there should be a local overabundance of
young NSs which can appear as hot cooling objects, as gamma-ray sources etc.

\begin{figure}[t] 
\epsfig{file=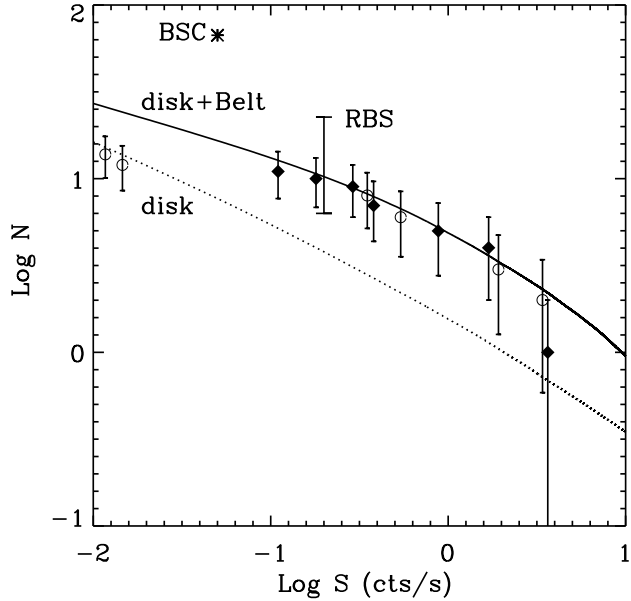, width=0.8\textwidth}
\caption{Log~N~--~Log~S distribution for close-by cooling INSs. 
Black symbols are plotted if the dimmest source at specified flux is
one of the "Magnificent seven". Otherwise we plot an opaque symbol
(see the list of sources in table~1).
Two lines represent results of calculation. Dotted line --- only stars
from the galactic disc contribute to the Log~N~--~Log~S distribution.
Solid line --- contribution of the Gould Belt is added. "RBS" and "BSC"
are two observational limits, obtained from the ROSAT data
(RBS: Schwope et al. 1999; BSC: Rutledge et al. 2003).
From Popov et al. (2003a).}
\label{lognlogs}
\end{figure}

Grenier (2003) estimated a number of possible unidentified EGRET sources
originated from the Belt. We (Popov et al. 2003a) calculated Log~N~--~Log~S
distribution of cooling NSs in the solar vicinity, which can be observed
by ROSAT and other X-ray missions. Results are shown in the
fig.~\ref{lognlogs}.

\subsection{Period evolution of INSs and accretion from the ISM}

 In early 70s it was suggested that INSs can be observed due to accretion of
the ISM and that significant part of INSs is at that stage. 
Actually, for such a prediction it is necessary to make some
assumptions about magnetorotational evolution of NSs. 
In particular, that a combination of spin period, magnetic field, spatial
velocity and density of the medium is such that accretion is possible
and proceeds at nearly Bondi rate.
Recent data shows that these assumptions were incorrect. ROSAT observations
resulted in just a few radioquiet INSs non of which is considered to be a
good candidate to be an accreting INSs. There are many reasons for that 
(see Popov et al. 2003b for discussion). 

 Recent studies show that it is very difficult for an INS to reach the stage
of accretion. There are three main reasons for that:\\
{\it i).} High spatial velocity.\\
{\it ii).} Magnetic field decay.\\
{\it iii).} Long subsonic propeller stage.

 Subsonic propeller stage was introduced by Davies and Pringle (1981).
Recently Ikhsanov (2003) re-investigated this issue in connection with INSs.
His main conclusion is that in a realistic situation an INS
spends a significant part  of its life ($ > 10^9$~yrs)
in this stage during which only a small amount of matter can diffuse inwards
and reach the star surface. If this situation is realized in nature than
even many of
low-velocity INSs may never become bright accretors. Long subsonic
propeller stage also leads to very long spin periods of INSs at the onset of
accretion, much longer, than it was suggested for example in Prokhorov et
al. (2002), who explored period evolution of INSs in some details.

Even if an INS starts to accrete, than its luminosity can be very low
due to:\\
{\it i).} Heating.\\
{\it ii).} Magnetospheric effects.\\
{\it iii).} Low accretion rate due to turbulence etc.\\

Heating was discussed in detailes by Blaes et al. (1995).
 During the last two years last two topics attracted much interest.
Magnetospheric effects were studied in a serie of papers by Romanova,
Toropina et al. 2D MHD simulations
have shown that accretion onto a rotating dipole is substantially different
from that onto an unmagnetized star (Romanova et al. 2003;
Toropina et al. 2003) Magnetic effects scales with the magnetic field
strength. For lower field they are less pronounced.

Recent investigations strongly support the idea that
the Bondi rate is just an upper limit which is rarely realized in nature 
(see Popov et al. 2003b and references therein).
2D and 3D simulations of accretion flows show that
convection and other effects
can reduce the accretion rate by orders of magnitude.
In particular Igumenshchev et al. (2002, 2003) explored so-called convection
dominated acretion flows (CDAF). 
Proga and Begelman (2003) stidied accretion with
low angular momentum. Both studies were done for black hole accretion
and showed very low accretion efficiency.
However, these results in principle can be applied to isolated NSs
(see Perna et al. 2003). 

According to all
these studies there is not much hope to observe accreting INSs.

\section{Conclusion. What do we -- astrophysicists -- want from 
QCD theorists \& Co.?}

Astronomy is in some sence a unique science:
we have only emission from objects under investigation.
Because of that there is a wide field for speculations.
Having a lot of uncertain parameters to explain properties
of observed compact objects
we have troubles even without quark stars! 

What do we want to have from theoretical physicists
as an input for astrophysical models of compact objects
to produce some output comparable with observations.
We want initial parameters of quark stars + evolutionary laws:

\begin{itemize}
\item Initial spin period, magnetic field, spatial velocity,
mass, radius etc.\\
(plus possible correlations between them).
\item "Ejectorability" (ability to produce relativistic wind,
i.e. to produce a radio pulsar).
\item Emission properties of the surface of bare strange stars.
\item Cooling curves.
\item Magnetic field decay.
\end{itemize}

We hope that this brief review will help to link advanced theoretical
research in physics of extremely dense matter with observational properties
of compact objects.
 

%

 
\begin{acknowledgments} 
SP thanks the Organizers for hospitality and financial support.
\end{acknowledgments}

\end{document}